\documentclass[prb,twocolumn]{revtex4}
\usepackage{graphicx}
\usepackage{dcolumn}
\usepackage{bm}
\usepackage[english]{babel}
\usepackage{color}

\begin{document}

\title{Direct biexciton generation in Si nanocrystal by a single photon}

\author{S.A. Fomichev, V.A. Burdov\footnote{Corresponding author, e-mail: vab3691@yahoo.com}}

\affiliation{Lobachevsky State University of Nizhny Novgorod, 23 Gagarin avenue, 603022 Nizhny Novgorod, Russian Federation}

\begin{abstract}

It has been shown theoretically that strong quantum confinement regime in Si nanocrystals promotes the highly efficient simultaneous excitation of two electron-hole pairs (biexciton) by a single photon. The rate (inverse lifetime) of biexciton generation has been calculated analytically as function of the nanocrystal radius. In contrast to the case of a nanocrystal formed of direct-band-gap semiconductor, the size-dependence of the rate in Si nanocrystal turns out to be sharp enough. At radii approaching a nanometer, the lifetime of biexciton generation falls into the nanosecond range.

\end{abstract}

\maketitle

\section{Introduction}

As known, interband optical transitions in silicon are, in fact, forbidden due to the indirect band gap. Because of a violation of the momentum conservation law, it is impossible to make a single-electron optical transition without any additional auxiliary mechanisms, such as, e.g., phonon assistance. In Si nanocrystals, this problem is partially overcome, since strict conservation of momentum during the transition is no longer required. This property manifests itself the more strongly the smaller the nanocrystal size. Nevertheless, the transition rate remains low enough even in small nanocrystals, since they largely inherit various features of the band structure of the bulk material.

Obviously, the momentum conservation law during the absorption of a single photon can be easily performed at a two-electron transition. If a photon transfers simultaneously two electrons from the vicinity of the valence band maximum to the vicinities of the two oppositely located in the Brillouin zone conduction band minima, no any ``additional'' particles, like, e.g., phonons, are required for this process, since the total two-electron momentum after the transition tends to zero similarly to that before the transition. The photon energy in this case should, at least slightly, doubly exceed the energy gap. Such a simultaneous two-electron transition is usually treated as a direct biexciton generation caused by a photon absorption in a nanocrystal.

Here, we examine a possibility of direct creating two electron-hole pairs, i.e. biexciton, by a single photon in a Si nanocrystal. We denote the valence single-particle initial states of the two electrons participating in the transition by the symbols $v$ and $v'$, while $c$ and $c'$ stand for their final single-particle states in the conduction band. Initially, a general expression for the radiative rate $\tau^{-1}$ of the biexciton creation will be obtained within the frames of a many-particle picture. Then, the single-particle wave functions from the envelope function approximation will be used for the transition matrix element and rate calculations, which allows us to determine the dependence of $\tau^{-1}$ on the nanocrystal radius $R$.

It should be noted that the process of the direct biexciton creation in nanocrystals was discussed earlier~\cite{Klimov3,Marri,Melnychuk} as applicable to the problem of carrier multiplication or multi-exciton generation. Various research groups were focused on the determination of the threshold energy and quantum efficiency of the process,~\cite{Klimov1,Klimov2,Beard,Gordi} population dynamics of biexciton states,~\cite{Hyeon,Deuk} and mechanisms of the biexciton creation.~\cite{Klimov1,Klimov2,Prezhdo,Shabaev} Rabani and Baer~\cite{Rabani} used the Green's function approach and calculated the densities of exciton and biexiton states in small CdSe, InAs and Si clusters and demonstrated that at energies exceeding twice the energy gap, the density of biexiton states increases sharply.

In the present paper we analytically calculate the rate of biexiton generation due to photon absorption in a silicon nanocrystal depending on the nanocrystal size. We show that with a decrease in the nanocrystal size, the rate of biexiton generation in a Si nanocrystal increases and at radii $R\sim 1$ nm becomes even comparable with the rates of single-exciton generation in nanocrystals of direct-band-gap semiconductors.

\section{Many-particle approach}

Initially, we have to define the Hamiltonian operator for our many-electron system. So far, we do not take into account interaction of the electron subsystem with electromagnetic field. Consequently, the Hamiltonian of the electron subsystem can be written as
\begin{equation}
\hat{\mathcal{H}} = \sum_{ik}E_{ik}\hat{a}_i^{+}\hat{a}_k + \frac{1}{2}\sum_{i,j,k,l}U_{ij}^{kl}\hat{a}_i^{+}\hat{a}_k^{+}\hat{a}_l\hat{a}_j,
\end{equation}
where the summation is carried out over the occupied states only,
\begin{equation}
E_{ik} = \langle\psi_i|\left(\frac{\hat{\bf p}^2}{2m} + V({\bf r})\right)|\psi_k\rangle
\end{equation}
is the matrix element of the electron kinetic and potential energies in the field of the crystal lattice $V({\bf r})$ calculated with respect to the Hartree-Fock single-electron wave functions $\psi_i(q)$ and $\psi_k(q)$, where $q = \{{\bf r};s_z\}$ is a combination of three-dimensional continuously varied position-vector ${\bf r}$ and discrete spin variable $s_z$, $m$ is the free electron mass, and operators  $\hat{a}_i^{+}$ and $\hat{a}_k$ create and annihilate electrons in $i$-th and $k$-th Hartree-Fock states, respectively. Matrix element of the Coulomb interaction potential energy
\begin{equation}
U({\bf r},{\bf r}') = \frac{e^2}{|{\bf r} - {\bf r}'|},
\end{equation}
is defined as
\begin{equation}
U_{ij}^{kl}=\int\psi_i^\ast(q)\psi_k^\ast(q')U({\bf r},{\bf r}')\psi_l(q')\psi_j(q)dqdq',
\end{equation}
Here, we use the basis of the single-particle Hartree-Fock states. Accordingly, it is convenient to rewrite the Hamiltonian operator in the spirit of the configuration interaction method taking the Hartree-Fock approximation as a zero-order one.

Then the Hamiltonian is represented by the sum of two parts:
\begin{equation}
\hat{\mathcal{H}} = \hat{\mathcal{H}}_0 + \hat{\mathcal{H}}_1,
\end{equation}
where $\hat{\mathcal{H}}_0$ is the diagonalized zero-order Hamiltonian
\begin{equation}
\hat{\mathcal{H}}_0 = \sum_{i}\varepsilon_i\hat{a}_i^{+}\hat{a}_i,
\end{equation}
where $\varepsilon_i$ differs from $E_{ik}$ by the matrix elements of the Hartree-Fock operator which allows to diagonalize the zero-order Hamiltonian:
\begin{equation}
\varepsilon_i = \langle\psi_i|\left(\frac{\hat{\bf p}^2}{2m} + V({\bf r})+ \hat{V}^{HF}\right)|\psi_i\rangle.
\end{equation}
Here,
\begin{equation}
\hat{V}^{HF} = V^{(H)}({\bf r}) + \hat{V}^{(X)}
\end{equation}
is the Hartree-Fock operator consisted of two parts: the Hartree potential energy
\begin{equation}
V^{(H)}({\bf r}) = \sum_i\int U({\bf r},{\bf r}')|\psi_i(q')|^2dq'
\end{equation}
and the non-local operator of the exchange interaction $\hat{V}^{(X)}$ that is defined as:
\begin{equation}
\hat{V}^{(X)}|\psi_j\rangle = \sum_i\int U({\bf r},{\bf r}')\psi_i^*(q')\psi_i(q)\psi_j(q')dq'.
\end{equation}
The second part
\begin{equation}
\hat{\mathcal{H}}_1 = \frac{1}{2}\sum_{i,j,k,l}U_{ij}^{kl}\hat{a}_i^{+}\hat{a}_k^{+}\hat{a}_l\hat{a}_j - \sum_{kj}V^{HF}_{kj}\hat{a}_k^{+}\hat{a}_j,
\end{equation}
where
\begin{eqnarray}
V^{HF}_{kj} & = & \langle\psi_k|\hat{V}^{HF}|\psi_j\rangle
\nonumber \\
& = & \sum_i\int\int\frac{\psi_k^*(q)\psi_i^*(q')e^2\psi_i(q')\psi_j(q)}{|{\bf r} - {\bf r}'|}dqdq'
\nonumber \\
& - & \sum_i\int\int\frac{\psi_k^*(q)\psi_i^*(q')e^2\psi_j(q')\psi_i(q)}{|{\bf r} - {\bf r}'|}dqdq'
\nonumber \\
& = & \sum_i(U_{kj}^{ii} - U_{ki}^{ij}).
\end{eqnarray}
We shall treat $\hat{\mathcal{H}}_1$ as a perturbation.

We assume the strong quantum confinement regime for the nanocrystal, i.e. suppose its radius to be much less than the effective exciton Bohr radius. This means that some typical size-quantization energy is much greater than a characteristic Coulomb energy in a nanocrystal. In this case it is indeed possible to treat the Coulomb interaction, defined by the Hamiltonian $\hat{\mathcal{H}}_1$, as a weak perturbation.

Consequently, we may search for wave functions of the many-electron system in the form of a series of perturbation theory. Restricting ourselves with the first order one can write the many-electron wave function in the form:
\begin{equation}
|\Psi_M\rangle = \left(1 + \sum_{J\neq M}\frac{|\Phi_J\rangle\langle\Phi_J|\hat{\mathcal{H}}_1}{\mathcal{E}_M - \mathcal{E}_J}\right)|\Phi_M\rangle,
\end{equation}
where $|\Phi_M\rangle$ and $\mathcal{E}_M$ are the wave function and the energy, respectively, of the many-electron $M$-th stationary state---the eigenstate and eigenvalue of the zero-order Hamiltonian $\hat{\mathcal{H}}_0$:
\begin{equation}
\hat{\mathcal{H}}_0|\Phi_M\rangle = \mathcal{E}_M|\Phi_M\rangle.
\end{equation}
$|\Phi_M\rangle$ is represented by the Slater determinant built of the single-particle wave functions $\psi_i^{(M)}$ corresponding to the $M$-th stationary state.

Now, we include into the consideration an interaction of the electron subsystem with the electromagnetic field that is described by the operator
\begin{equation}
\hat{W} = \sum{\bf v}\hat{{\bf p}} = \sum_{ik}{\bf v}{\bf p}_{ik}\hat{a}_i^{+}\hat{a}_k,
\end{equation}
where
\begin{equation}
{\bf v} = \frac{eA}{mc}{\bf e}_{\alpha},
\end{equation}
$e$ is an absolute value of the electron charge, $A$ is the vector-potential amplitude, $c$ is the speed of light, ${\bf e}_{\alpha}$ is the polarization vector, and the momentum matrix element is calculated with respect to the single-electron wave functions: ${\bf p}_{ik} = \langle\psi_i|\hat{{\bf p}}|\psi_k\rangle$. As usual, we suppose the wave vector of the electromagnetic wave tending to zero. Therefore, vector ${\bf v}$ does not depend on the electron position-vector. The first sum is the sum over particles, while $i$ and $k$ in the second sum run over the filled Hartree-Fock states.

The electron subsystem can change its state under the action of the electromagnetic wave. Below, we consider the transition of the electron subsystem, initiated by a single photon, from its ground state, when all electrons completely populate the valence band, to the excited state, when two electrons leave two single-particle valence states $v$ and $v'$, and occupy two single-particle conduction states $c$ and $c'$. We call these stationary many-electron states ``initial'' and ``final'', respectively, and denote their wave functions as $|\Psi_I\rangle$ and $|\Psi_F\rangle$.

\section{Matrix element of the transition}

The zero-order wave function of the ground state of the electron subsystem can be symbolically represented by the ket-vector
\begin{equation}
|\Phi_I\rangle = |111...11_v1_{v'}\parallel0_c0_{c'}00...\rangle,
\end{equation}
where double vertical line means the energy gap, and energy of the states rises from left to right. The zero-order excited state is defined as
\begin{equation}
|\Phi_F\rangle = \hat{a}_c^{+}\hat{a}_{c'}^{+}\hat{a}_v\hat{a}_{v'}|\Phi_I\rangle.
\end{equation}
It should be noted here that, strictly speaking, the single-particle Hartree-Fock wave functions, from which the Slater determinants are built, differ for the initial and final states, because both Hartree and exchange operators slightly change due to the replacement of two terms ($\psi_v$ and $\psi_{v'}$) by two other terms ($\psi_c$ and $\psi_{c'}$) in the sums defining $V^{(H)}({\bf r})$ and $\hat{V}^{(X)}$. We, however, neglect this difference in the Hartree-Fock wave functions assuming it insignificant.

After some transformations of Eq. (13) the wave functions of the ground and excited states up to the first-order corrections take the form:
\begin{equation}
|\Psi_I\rangle = \left(1 - \sum_{J\neq I}\frac{|\Phi_J\rangle\langle\Phi_J|\hat{\mathcal{H}}_1}{\mathcal{E}_J}\right)|\Phi_I\rangle,
\end{equation}
and
\begin{equation}
|\Psi_F\rangle = \left(1 + \sum_{J\neq F}\frac{|\Phi_J\rangle\langle\Phi_J|\hat{\mathcal{H}}_1}{\mathcal{E}_F - \mathcal{E}_J}\right)|\Phi_F\rangle.
\end{equation}
Here, we took into account that the initial state of the electron subsystem is its ground state. For convenience, we set $\mathcal{E}_I = 0$, so that the final state has the energy $\mathcal{E}_F = \varepsilon_c + \varepsilon_{c'} - \varepsilon_v - \varepsilon_{v'}$. In a nanocrystal, the ground state energies are degenerate in both the valence and conduction bands. Therefore, $\varepsilon_v = \varepsilon_{v'} = \varepsilon_{v0}$ and $\varepsilon_c = \varepsilon_{c'} = \varepsilon_{c0}$, where $\varepsilon_{v0}$ and $\varepsilon_{c0}$ are the ground state energies in the valence and conduction bands, respectively. As a result, $\mathcal{E}_F = 2\varepsilon_g$, where $\varepsilon_g$ is a single-particle gap of the nanocrystal.

Obviously, the expression for $W_{FI}$ reduces to
\begin{eqnarray}
W_{FI} && = \sum_{ik}{\bf v}{\bf p}_{ik}\sum_{J\neq F}\frac{\mathcal{H}^{(1)}_{FJ}}{\mathcal{E}_F - \mathcal{E}_J}\langle\Phi_J|\hat{a}_i^{+}\hat{a}_k|\Phi_I\rangle
\nonumber \\
&& - \sum_{ik}{\bf v}{\bf p}_{ik}\sum_{J\neq I}\frac{\mathcal{H}^{(1)}_{JI}}{\mathcal{E}_J}\langle\Phi_F|\hat{a}_i^{+}\hat{a}_k|\Phi_J\rangle,
\end{eqnarray}
where $\mathcal{H}^{(1)}_{KM} = \langle\Phi_K|\hat{\mathcal{H}}_1|\Phi_M\rangle$. The first term in $W_{FI}$ is caused by the corrections to the final state of the system [Eq. (20)], while the second one appears due to the corrections to the initial state [Eq. (19)]. As can be seen, matrix element $\langle\Phi_J|\hat{a}_i^{+}\hat{a}_k|\Phi_I\rangle$ in the first sum is different from zero only if $J$-th state is a single-exciton one, since the initial (ground) state has no excitons at all, and the pair operator $\hat{a}_i^{+}\hat{a}_k$ may increase (or decrease) the number of excitons by only one. Similarly, taking into account that the final state has two excitons, one can conclude that in the second sum, matrix element $\langle\Phi_F|\hat{a}_i^{+}\hat{a}_k|\Phi_J\rangle$ formally allows $J$-th state with one, two, or three excitons. However matrix element $\mathcal{H}^{(1)}_{JI} \neq 0$ only for single- and biexciton states. Moreover, as our calculations show, only biexciton states contribute to the second sum. Thus, the corrections to the final state describe the two-electron excitation through virtual exciton states, while the corrections to the initial state induce the two-electron transitions through virtual biexciton states. Both kinds of these transitions were discussed earlier as possible ways for obtaining multi-exciton generation in A$_4$B$_6$ and A$_2$B$_6$~\cite{Melnychuk,Klimov1,Klimov2,Prezhdo} or Si~\cite{Klimov3,Marri,Rabani} nanocrystals.

It is seen also that $W_{FI}$ is directly defined by the matrix elements of the operator $\hat{\mathcal{H}}_1$ which, in fact, describes correlation effects in the electron gas. Operator $\hat{\mathcal{H}}_1$, according to Eqs (11) and (12), is a linear combination of various Coulomb matrix elements $U_{nn'}^{mm'}$. This means that simultaneous inter-band transition of two electrons caused by the absorption of only one photon can be treated as a result of correlations in the electron subsystem. Evidently, if we do not take an interaction between the electrons into account, all matrix elements $U_{nn'}^{mm'} = 0$. As a consequence, matrix element $W_{FI}$ and the probability of such a transition also turn into zero.

Using Eqs (17), (18), and (21) it is possible to obtain after some algebra following expression for $W_{FI}$:
\begin{eqnarray}
&& W_{FI} = {\bf v}\sum_n\frac{{\bf p}_{nv}(U_{c'n}^{cv'} - U_{cn}^{c'v'}) + {\bf p}_{nv'}(U_{cn}^{c'v} - U_{c'n}^{cv})}{2\varepsilon_g + \varepsilon_{v0} - \varepsilon_n}
\nonumber \\
&& + {\bf v}\sum_n\frac{{\bf p}_{cn}(U_{nv}^{c'v'} - U_{nv'}^{c'v}) + {\bf p}_{c'n}(U_{nv'}^{cv} - U_{nv}^{cv'})}{2\varepsilon_g - \varepsilon_{c0} + \varepsilon_n}.
\end{eqnarray}
Here, the first sum describes the contribution to $W_{FI}$ from the virtual transitions through exciton states, while the second sum is due to the virtual transitions through biexciton states. As seen, if the initial states $v$ and $v'$, or the final states $c$ and $c'$, coincide (we mean spatial parts of their wave functions), then $W_{FI} = 0$. Thus, the initial states, as well as the final states, should be different to provide nonzero matrix element of the transition.

\section{Total rate of the biexciton creation}

In order to calculate the rate of the direct biexciton generetion we use the Fermi golden rule:
\begin{equation}
\tau^{-1} = 4\times \frac{2\pi}{\hbar}\sum_{{\bf q},\alpha}\sum_{I,F}|W_{FI}|^2\delta(\hbar\omega - 2\varepsilon_g),
\end{equation}
where ${\bf q}$ and $\alpha$ stand for the photon wave vector and polarization, respectively, and the factor ``4'' arises due to summation over spins of two particles. Accordingly, we have to calculate $W_{FI}$, i.e. matrix elements of the momentum operator as well as of the Coulomb potential energy, as it follows from Eq. (22).

For this purpose the Hartree-Fock wave functions are necessary. It is quite clear, however, that solving the Hartree-Fock equations is a very time-consuming and cumbrous procedure. Therefore, here, instead of the exact solutions of the Hartree-Fock equations, we shall use some approximate expressions for the single-particle wave functions obtained within the envelope function approximation. Accordingly, the wave functions will be the products of smooth envelope functions and sharply varied Bloch functions. Within this approach, the pattern of electronic excitation in a nanocrystal inherits to a large extent the features of this process in bulk silicon. On this reason we frequently use here bulklike terminology and interpretation.

The initial single-electron states belong to the top of the valence band, and correspond to the $\Gamma$-point in the Brillouin zone. As known, the $\Gamma$-point in the valence band is triply degenerate, and three periodic Bloch functions $u_1$, $u_2$, and $u_3$ in a diamond structure transform according to the irreducible representation $\Gamma'_{25}$ of the wave-vector group. The valence band in silicon is anisotropic, and determination of the envelope functions for the valence states is not a simple problem. However, for the ground states, an isotropic approximation may be employed.~\cite{JETP,PRB07} We average the $kp$-matrix 3$\times$3 over angles in the ${\bf k}$-space and obtain isotropic and diagonal matrix with triply degenerate eigenvalue
\begin{equation}
\varepsilon_{v0} = - \frac{\hbar^2\pi^2}{2m_vR^2},
\end{equation}
where $m_v = 3m_0/(L + 2M) = 0.24m$ is a mean effective mass in the valence band, and $L = 5.8$, $M=3.4$ for silicon.~\cite{Landolt} Consequently, the single-particle wave functions of the initial states (which coincide with the ground states in our case) can be written as follows:
\begin{eqnarray}
\psi_v({\bf r}) &=& \psi_{v_i}({\bf r}) = \phi(r)u_i({\bf r}),
\nonumber \\
\psi_{v'}({\bf r}) &=& \psi_{v_k}({\bf r}) = \phi(r)u_k({\bf r}),
\end{eqnarray}
where $i \ne k$. Here, $\phi(r)$ stands for the envelope function
\begin{equation}
\phi(r) = \frac{\sin(\pi r/R)}{r\sqrt{2\pi R}}
\end{equation}
for the ground valence electron states in spherical nanocrystal of radius $R$.

The final single-electron states coincide with two of six ground states in the conduction band (spin is not taken into account). The wave functions of the final single-electron states are formed according to the tetrahedral point symmetry of the nanocrystal as the product of $\phi(r)$ and some linear combinations of the Bloch functions of the six energy minima in the conduction band. These linear combinations should be basis functions of three irreducible representations (one-dimensional unit representation $A_1$, two-dimensional $E$, and three-dimensional $T_2$) of the point group $T_d$.~\cite{KL1,KL2} As a result, the wave function of each the final single-particle state, $\psi_c$ or $\psi_{c'}$, can be one of the following six functions:
\begin{eqnarray}
&& \psi_A({\bf r}) = \phi(r)\frac{\psi_{k_x} + \psi_{k_x}^* + \psi_{k_y} + \psi_{k_y}^* + \psi_{k_z} + \psi_{k_z}^*}{\sqrt{6}}, \nonumber \\
&& \psi_E^{(1)}({\bf r}) = \phi(r)\frac{\psi_{k_x} + \psi_{k_x}^* - \psi_{k_y} - \psi_{k_y}^*}{2}, \nonumber \\
&& \psi_E^{(2)}({\bf r}) = \phi(r)\frac{\psi_{k_x} + \psi_{k_x}^* + \psi_{k_y} + \psi_{k_y}^* - 2\psi_{k_z} - 2\psi_{k_z}^*}{\sqrt{12}}, \nonumber \\
&& \psi_T^{(1)}({\bf r}) = \phi(r)\frac{\psi_{k_x} - \psi_{k_x}^*}{\sqrt{2}}, \nonumber \\
&& \psi_T^{(2)}({\bf r}) = \phi(r)\frac{\psi_{k_y} - \psi_{k_y}^*}{\sqrt{2}}, \nonumber \\
&& \psi_T^{(3)}({\bf r}) = \phi(r)\frac{\psi_{k_z} - \psi_{k_z}^*}{\sqrt{2}},
\end{eqnarray}
where $\psi_{k_l} = u_{k_l}({\bf r})\exp\{ikx_l\}$, $k = 0.85\times(2\pi/a)$ is the distance from the center of the Brillouin zone to any of the six energy minima in the conduction band, and $a$ is the lattice constant. Functions $u_{k_l}$ transform according to irreducible representation $\Delta_1$ of the wave-vector group in a diamond structure.

Similarly to what was done in the valence band, we average the dispersion law in the conduction-band valleys over angles in the ${\bf k}$-space, which yields six-fold degenerate energy level of the ground state
\begin{equation}
\varepsilon_{c0} = E_g + \frac{\hbar^2\pi^2}{2m_cR^2},
\end{equation}
where $E_g = 1.17$eV is the energy gap of bulk Si, $m_c = 3m_lm_t/(2m_l + m_t) = 0.26m$ is an isotropic effective mass in the conduction band, and $m_l$ and $m_t$ stand for the ``longitudional'' and ``transverse'' effective masses, respectively.

Hereafter, we use the model of almost free electrons for the Bloch amplitudes in all bands in order to perform quantitative estimations of the transition rates. Within this model
\begin{eqnarray}
u_n({\bf r}) & = & 2\cos\frac{2\pi x_l}{a}\cos\frac{2\pi x_m}{a}\sin\frac{2\pi x_n}{a} \nonumber \\
& + & 2\sin\frac{2\pi x_l}{a}\sin\frac{2\pi x_m}{a}\cos\frac{2\pi x_n}{a},
\end{eqnarray}
\begin{eqnarray}
u_{k_n}({\bf r}) &=& \sqrt{2}\cos\frac{2\pi x_l}{a}\cos\frac{2\pi x_m}{a}e^{-2\pi ix_n/a} \nonumber \\
&+& i\sqrt{2}\sin\frac{2\pi x_l}{a}\sin\frac{2\pi x_m}{a}e^{-2\pi ix_n/a},
\end{eqnarray}
where each of indices $l$, $m$, $n$ runs over values $x$, $y$, and $z$ (or 1, 2, and 3) and all are different in these expressions.

It is well known that within the model of almost free electrons the $\Gamma$-point (center of the Brillouin zone) formed of the wave vectors of [111]-family (in units of $2\pi/a$) is eight-fold degenerate. Eight plane waves form the basis of some reducible representation of the group of the $\Gamma$-point wave vector. Periodic lattice potential splits the eight-fold degenerate energy level into four levels---two non-degenerate, and two triply-degenerate corresponding to the irreducible representations $\Gamma_1$ (one-dimensional), $\Gamma'_2$ (one-dimensional), $\Gamma_{15}$ (three-dimensional), and $\Gamma'_{25}$ (three-dimensional). $|\Gamma'_{25}\rangle$ states form the top of the valence band, while $|\Gamma_1\rangle$, $|\Gamma'_2\rangle$, and $|\Gamma_{15}\rangle$ states form in the $\Gamma$-point three energy branches in the conduction band. All other energy bands in the $\Gamma$-point are built from the reciprocal lattice vectors different from the vectors of the family [111].

Among intermediate states $n$, as it follows from the model of almost free electrons and the symmetry of the Bloch functions, only the states of the conduction bands $\Gamma'_2$ and $\Gamma_{15}$, as well as the states of the valence band $\Delta_5$, contribute to the momentum matrix elements and, consequently, to $W_{FI}$, as shown in Fig. 1 by the arrows depicting the corresponding virtual transitions.

\begin{figure}[t]
  \centering
  \includegraphics[scale=1]{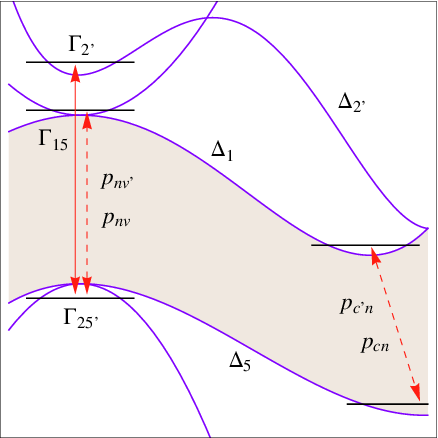}
  \caption{Allowed electron virtual transitions to the intermediate states $\Gamma'_2$, $\Gamma_{15}$ of the conduction band, and $\Delta_5$-state of the valence band. Dashed arrows indicate a relative weakness of the transitions. In the case of $c\leftrightarrow n$ and $c'\leftrightarrow n$ transitions such a weakness is caused by their slightly indirect character that is shown with tilted arrow.} \label{figure1}
\end{figure}

Within the framework of the envelope function approximation matrix elements ${\bf p}_{nv}$ and ${\bf p}_{nv'}$ can be written as
\begin{equation}
{\bf p}_{nv} = {\bf p}_{nv}^{(0)}\langle \phi_n|\phi\rangle, \ \ {\bf p}_{nv'} = {\bf p}_{nv'}^{(0)}\langle \phi_n|\phi\rangle,
\end{equation}
where ${\bf p}_{nv}^{(0)}$ and ${\bf p}_{nv'}^{(0)}$ are bulk-like matrix elements, while $\phi_n$ stands for the envelope function of the $n$-th intermediate state in the conduction band.

$\Gamma'_2$ is a simple band with a positive isotropic effective mass. Therefore the ground state in this band has the wave function with $\phi_n(r) = \phi(r)$:
\begin{equation}
\psi_n({\bf r}) = \phi(r)u_{\Gamma'_2}({\bf r}),
\end{equation}
where
\begin{eqnarray}
u_{\Gamma'_2}({\bf r}) & = & 2\cos\frac{2\pi x}{a}\cos\frac{2\pi y}{a}\cos\frac{2\pi z}{a}
\nonumber \\
& - & 2\sin\frac{2\pi x}{a}\sin\frac{2\pi y}{a}\sin\frac{2\pi z}{a}.
\end{eqnarray}
As a result, $\langle \phi_n|\phi\rangle = 1$, and momentum matrix elements ${\bf p}_{nv}$ and ${\bf p}_{nv'}$ for $\Gamma'_2$ band are reduced to their bulk-like values:
\begin{equation}
\langle u_{\Gamma'_2}|\hat{p}_j|u_j\rangle = -i\frac{2\pi\hbar}{a}.
\end{equation}

The band $\Gamma_{15}$ is anisotropic and triply degenerate. This means that $\phi_n(r)$ for the $\Gamma_{15}$ band should be represented by a linear combination of various functions of $s$, $p$, $d$, \textit{etc.} types. Only $1s$-function contributes to the matrix element, so that $\langle \phi_n|\phi\rangle < 1$ in this case. This decreases the values of ${\bf p}_{nv}$ and ${\bf p}_{nv'}$ for the $\Gamma_{15}$ band. Moreover, as our calculations show, the Coulomb matrix elements with $n = \Gamma_{15}$ also turn out to be small compared to those with $n = \Gamma'_2$. Therefore, in the following we neglect the contribution from the $\Gamma_{15}$ band.

In the cases, when the $n$-th Bloch function transforms according to the irreducible representation $\Gamma_1$ in the conduction band, or $\Gamma'_{25}$ in the valence band, the matrix elements ${\bf p}_{nv} = {\bf p}_{nv'} = 0$ due to symmetry. All other states, whose Bloch functions are built of the plane waves with the wave vectors different from [111], i.e. all states from all other energy bands, also yield ${\bf p}_{nv} = {\bf p}_{nv'} = 0$. This happens precisely because the wave vectors of the plane waves, from which the functions of all other bands are constructed, differ from [111] by the vectors of the reciprocal lattice and at the same time have different from [111] absolute values. It is, of course, the consequence of using here the free-electron model.

Matrix elements ${\bf p}_{cn}$ and ${\bf p}_{c'n}$ for the $n$-th Bloch states belonging to the valence band $\Delta_5$ can be calculated similarly to ${\bf p}_{nv}$ and ${\bf p}_{nv'}$. However we should take into account that the wave function of the $n$-th state in this case has the form $\psi_n({\bf r}) = \phi(r)\sum_iC_{ni}\varphi_i({\bf r})$, where $\varphi_i({\bf r})$ is the Bloch function at the $i$-th $X$-point, and $C_{ni}$ stand for some coefficients determined by the tetrahedral symmetry of the system. Such a form arises because minimum of the valence $\Delta_5$ energy branch is located at the $X$-point in contrast to the conduction $\Delta_1$ energy branch which has its minimum at the distance $\kappa = 0.15\times 2\pi/a$ from the $X$-point. Thus, virtual transition from $\Delta_1$ to $\Delta_5$ turns out to be slightly indirect, which is shown in Fig. 1 by the tilted arrow. As a result, ${\bf p}_{cn}$ and ${\bf p}_{c'n}$ will be proportional to the matrix elements $\langle\phi|\exp(\pm i\kappa x_l)|\phi\rangle$ which essentially decreases ${\bf p}_{cn}$ and ${\bf p}_{c'n}$---their values turn out to be order-of-magnitude less than ${\bf p}_{nv}$ and ${\bf p}_{nv'}$ [Eq. (34)], while for all other bands ${\bf p}_{cn} = {\bf p}_{c'n} = 0$ on the same reason as for the $\Gamma$-bands.

Consequently, in the subsequent calculations we neglect ${\bf p}_{cn}$ and ${\bf p}_{c'n}$ and omit the second sum in Eq. (22). This means that we do not take into account the contribution of the two-electron transitions through virtual biexciton states induced by the initial-state correction [Eq. (19)]. Similar approach was used earlier~\cite{Rabani} when considering carrier multiplication in silicon nanocrystals. In the first sum we restrict the summation only by the intermediate states of the $\Gamma'_2$ band, which yields:
\begin{equation}
W_{FI} = w(v,v') - w(v',v),
\end{equation}
where $w(a,b) = {\bf v}{\bf p}_{\Gamma'_2a}(U_{c'\Gamma'_2}^{cb} - U_{c\Gamma'_2}^{c'b})/\varepsilon$, and $\varepsilon = 2\varepsilon_g + \varepsilon_{v0} - \varepsilon_c(\Gamma'_2)$. Below, we consider various kinds of the initial and final states and focus on calculations of the Coulomb matrix elements in these cases.

Generally speaking, the initial states are always described by the wave functions [Eq. (25)] which transform according to the irreducible representation $T_2$ of the tetrahedral point group. Meanwhile, the wave functions of the final states, as was already pointed out above, can be basis functions of three irreducible representations ($A_1$, $E$, and $T_2$). Therefore, it is sufficient to consider various kinds of the final states only, since the initial states remain invariable.

1. The wave functions of both final states are basis functions of $T_2$ irreducible representation: $\psi_c = \psi_T^{(m)}$; $\psi_{c'} = \psi_T^{(l)}$ ($m \neq l$). In this case the wave functions are located in different areas in the ${\bf k}$-space, and their overlap is very weak. All the Coulomb matrix elements are negligibly small.

2. The wave functions of both final states are basis functions of $E$ irreducible representation: $\psi_c = \psi_E^{(1)}$; $\psi_{c'} = \psi_E^{(2)}$. In this case all the Coulomb matrix elements turn into zero due to symmetry.

3. The wave functions of the final states are basis functions of $A_1$ and $E$ irreducible representations: $\psi_c = \psi_A$; $\psi_{c'} = \psi_E^{(i)}$. In this case all the Coulomb matrix elements turn into zero due to symmetry.

4. The wave functions of the final states are basis functions of $T_2$ and $E$ irreducible representations. In this case ten Coulomb matrix elements have nonzero values:
\begin{eqnarray}
&& U_{T_2^{(1)}\Gamma'_2}^{E^{(1)}v_1} = - U_{E^{(1)}\Gamma'_2}^{T_2^{(1)}v_1} = U_{E^{(1)}\Gamma'_2}^{T_2^{(2)}v_2} = - U_{T_2^{(2)}\Gamma'_2}^{E^{(1)}v_2} = i\sqrt{3}U_0, \nonumber \\
&& U_{T_2^{(1)}\Gamma'_2}^{E^{(2)}v_1} = - U_{E^{(2)}\Gamma'_2}^{T_2^{(1)}v_1} = U_{T_2^{(2)}\Gamma'_2}^{E^{(2)}v_2} = - U_{E^{(2)}\Gamma'_2}^{T_2^{(2)}v_2} = iU_0, \nonumber \\
&& U_{E^{(2)}\Gamma'_2}^{T_2^{(3)}v_3} = - U_{T_2^{(3)}\Gamma'_2}^{E^{(2)}v_3} = 2iU_0,
\end{eqnarray}
where $U_0$ is defined by
$$
\frac{ie^2}{\sqrt{6}}\int d{\bf r}d{\bf r}'\frac{\phi^2(r)\phi^2(r')\psi^*_{k_z}({\bf r})u_z({\bf r})\psi_{k_z}({\bf r}')u_{\Gamma'_2}({\bf r}')}{|{\bf r} - {\bf r}'|},
$$
and can be represented as
$$
\frac{ie^2}{2\sqrt{6}\pi^2R^2}\sum_{{\bf g}}A_{\bf g}B^*_{\bf g}\int d{\bf p}\frac{F(pR)}{p^2({\bf b}_{\bf g} + {\bf k} - {\bf p})^2},
$$
where
$$
F(x) = \left(\text{Si}(x) - \frac{\text{Si}(x + 2\pi)}{2} - \frac{\text{Si}(x - 2\pi)}{2}\right)^2,
$$
$\text{Si}(x)$ stands for the sin-integral function, ${\bf b}_{\bf g}$ is the reciprocal lattice vector, ${\bf k} = k{\bf e}_z$, and $A_{\bf g}$ and $B_{\bf g}$ are the expansion coefficients of the products $u_{k_z}u_{\Gamma'_2}$ and $u_{k_z}u_z$, respectively, into the Fourier series. $F(pR)$ as a function of $p$ has a narrow peak of the width $\sim \pi/R$, which allows one to write approximately:
$$
\int d{\bf p}\frac{F(pR)}{p^2({\bf b}_{\bf g} + {\bf k} - {\bf p})^2} \approx \frac{4\pi S}{R({\bf b}_{\bf g} + {\bf k})^2}.
$$
Here, we have introduced
$$
S = \int\limits_{0}^{\infty}F(x)dx \approx 13.3.
$$
Coefficients $A_{\bf g}$ and $B_{\bf g}$ can be easily found from Eqs (29), (30), and (33), which yields
$$
\sum_{{\bf g}}\frac{A_{\bf g}B^*_{\bf g}}{({\bf b}_{\bf g} + {\bf k})^2} = - \frac{iC}{2}\left(\frac{a}{2\pi}\right)^2,
$$
where
$$
C = \frac{1}{1.15^2} - \frac{1}{4 + 0.85^2} - \frac{1/4}{8 + 0.85^2} + \frac{1/4}{8 + 1.15^2} \approx 0.54.
$$
As a result, one obtains
\begin{equation}
U_0 = \frac{CSe^2a^2}{4\sqrt{6}\pi^3R^3}.
\end{equation}

\begin{figure}[t]
  \centering
  \includegraphics[scale=1]{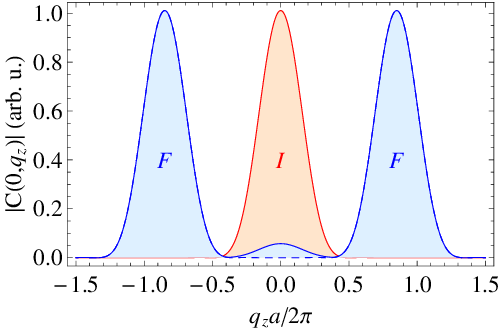}
  \caption{Absolute values of the two-electron wave functions of the initial ($I$) and final ($F$) states in the (${\bf Q},{\bf q}$)-representation at ${\bf Q}=0$ normalized to their maximum values as functions of dimensionless $q_z$. For the final state the wave functions are plotted for the cases of noninteracting (dashed line) and interacting (solid line) electrons.} \label{figure2}
\end{figure}

5. The wave functions of the final states are basis functions of $A_1$ and $T_2$ irreducible representations. In this case six Coulomb matrix elements have nonzero values:
\begin{eqnarray}
&& U_{T_2^{(1)}\Gamma'_2}^{A_1v_1} = U_{T_2^{(2)}\Gamma'_2}^{A_1v_2} = U_{T_2^{(3)}\Gamma'_2}^{A_1v_3} = i\sqrt{2}U_0, \nonumber \\
&& U_{A_1\Gamma'_2}^{T_2^{(1)}v_1} = U_{A_1\Gamma'_2}^{T_2^{(2)}v_2} = U_{A_1\Gamma'_2}^{T_2^{(3)}v_3} = - i\sqrt{2}U_0.
\end{eqnarray}

It is worth noting here that, within the frames of a less rigorous two-particle description of the considered system, an existence of these nonzero Coulomb matrix elements means that the two-electron wave function of the final state is not a simple product of two single-electron functions, e.g., $\psi_A\psi_T^{(3)}$, but also has a correction due to the inter-particle Coulomb interaction. In particular, we can write (neglecting the exchange):
\begin{eqnarray}
\Psi_F(1,2) &=& \psi_A(1)\psi_T^{(3)}(2) \nonumber \\
&+& i\frac{\sqrt{2}U_0}{\varepsilon(R)}\phi(1)\phi(2)u_z(1)u_{\Gamma'_2}(2).
\end{eqnarray}
At the same time, the two-electron wave function of the initial state remains the product of single-electron functions, e.g.:
\begin{equation}
\Psi_I(1,2) = \phi(1)\phi(2)u_z(1)u_x(2).
\end{equation}

It is now possible to introduce the total and relative two-electron wave-vectors ${\bf Q} = {\bf k} + {\bf k}'$ and ${\bf q} = ({\bf k} - {\bf k}')/2$, respectively, for these initial and final two-electron states and make the Fourier transform:
\begin{equation}
C({\bf Q},{\bf q}) = \int d{\bf r}_1d{\bf r}_2\Psi(1,2)e^{-i(\frac{{\bf Q}}{2}+{\bf q}){\bf r}_1}e^{-i(\frac{{\bf Q}}{2}-{\bf q}){\bf r}_2}.
\end{equation}
Obviously, $|C({\bf Q},{\bf q})|$ has a maximum at ${\bf Q} = 0$ for both initial and final states, which is, in fact, a consequence of the momentum conservation law for the two electrons. Meanwhile, the dependence on ${\bf q}$ strongly differs for the initial and final states. In Fig. 2 we have plotted $|C({\bf Q},{\bf q})|$ along the $q_z$ direction at ${\bf Q} = 0$. As seen, the initial state is located around $q_z = 0$, while the final state is mainly concentrated near the points $q_z = \pm k = \pm 0.85\times 2\pi/a$. If the electrons in the final state are treated as non-interacting ($U_0 = 0$ in Eq. (39), shown with dashed line in the figure), an overlap of $C_I(0,q_z)$ and $C_F(0,q_z)$ is very weak, and the process of biexciton creation turns out to be forbidden. If the Coulomb interaction is taken into account (solid line), then an extra peak at $q_z = 0$ appears on the dependence $|C_F(0,q_z)|$, which provides an overlap of the wave functions of the initial and final states in the reciprocal space. One can conclude that a participation of two electrons in the process allows to conserve the momentum during the transition, while the Coulomb interaction modifies the two-electron wave functions to provide their overlap in the ${\bf k}$-space.

We can now finalize the calculation of the biexciton generation rate. As was pointed out above, 16 various combinations of the initial and final states contribute to $W_{FI}$. The sum over $I$ and $F$ in Eq. (23) yields:
$\sum_{I,F}|W_{FI}|^2 = 48U_0^2(2\pi\hbar/a)^2|{\bf v}|^2/\varepsilon^2$. Since $|{\bf e}_\alpha|^2 = 1$, this sum does not depend on the photon polarization, which allows one to replace $\sum_\alpha \rightarrow 2$ in Eq. (23). Consequently, one obtains:
\begin{equation}
\tau^{-1} = \frac{256C^2S^2\sqrt{\epsilon_0}n(\omega_g)\kappa(\omega_g)\varepsilon_ge^6a^2}{\pi^4m^2c^3{\varepsilon}^2R^6}.
\end{equation}
Here, $\omega_g = 2\varepsilon_g/\hbar$ is the transition frequency, $n(\omega_g)$ is the number of photons with frequency $\omega_g$ in the electromagnetic resonator, and
\begin{equation}
\kappa(\omega) = \left(\frac{3\epsilon_0}{2\epsilon_0 + \epsilon(\omega)}\right)^2,
\end{equation}
where $\epsilon(\omega)$ and $\epsilon_0$ stand for permittivities of the nanocrystal and surrounding matrix, respectively. Parameter $\kappa$ is a squared local field factor that describes weakening the electromagnetic field inside the silicon nanocrystal related to that outside it.~\cite{Gibbs} We suppose the surrounding matrix to be a wide-band dielectric like SiO$_2$ whose energy gap is significantly wider than the photon energy. Therefore, it is possible to use static permittivity for this material. On the contrary, the nanocrystal permittivity should be function of $\omega$.

As seen, the rate has sufficiently sharp size dependence that is mainly defined by $R^{-6}$, but also by $\varepsilon_g$ and $\varepsilon$. Within the model of infinitely deep potential well, that was used above to describe electronic states in the nanocrystal, $\varepsilon$ turns into zero at $R$ a little greater than 1 nm. This leads to an unlimited rise of the rate. In order to avoid such an unphysical result, we employ here more accurate model of finite barriers and effective mass discontinuity when calculate $\varepsilon_g$ and $\varepsilon$. In particular, the barrier heights were chosen equal to 3.2 eV and 4.5 eV for the electrons and holes, respectively, which is typical for the contact of Si and amorphous SiO$_2$,~\cite{Wilk} while the energy and the effective mass in the $\Gamma'_2$ band  were set $E(\Gamma'_2) = 4.2$eV,~\cite{Landolt} and $m(\Gamma'_2) = 0.16m$,~\cite{Cardona} respectively. Both electron and hole effective masses outside the nanocrystal were chosen equal to the free electron mass $m$.

The parameters $n(\omega)$ and $\kappa(\omega)$ are not well known. Therefore, instead of the rate of the two-electron transition we calculate the product $(\tau\kappa(\omega_g)n(\omega_g))^{-1}$ and show its dependence on $R$ in Fig. 3 with thick line. In fact, it is possible to say that we calculate $\tau^{-1}$ at $\kappa(\omega_g)n(\omega_g) = 1$. It is suitable to compare this result for the biexciton radiative creation with results obtained earlier for exciton radiative creation in some similar nanocrystalline systems at the same radii. In particular, we consider here the single-electron radiative transitions in a: nanocrystal of direct-band-gap semiconductor; P-doped Si nanocrystal; undoped Si nanocrystal, where the transition is accompanied by the phonon assistance; and Si nanocrystal with chemically modified surface. The rate of a single-electron radiative transition is determined by
\begin{equation}
\tau^{-1} = \frac{4e^2\varepsilon_g|{\bf p}_{if}|^2n(\omega_g/2)\kappa(\omega_g/2)\sqrt{\epsilon_0}}{3m^2\hbar^2c^3}.
\end{equation}
where ${\bf p}_{if}$ denotes the momentum matrix element calculated with respect to the wave functions of the initial and final states.

\begin{figure}[t]
  \centering
  \includegraphics[scale=0.9]{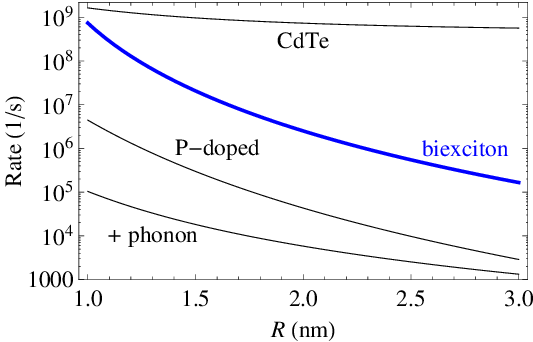}
  \caption{Rates of (from top to bottom): exciton creation in colloidal zinc-blende CdTe nanocrystal; biexciton creation in Si nanocrystal (thick line); exciton creation in P-doped Si nanocrystal; and phonon-assisted exciton creation in undoped Si nanocrystal.} \label{figure3}
\end{figure}

When we deal with the nanocrystal of direct-band-gap material, ${\bf p}_{if}$, in fact, coincides with the Kane's matrix element whose absolute value can be estimated from the model of almost free electrons, which yields: $|{\bf p}_{if}| = 2\pi\hbar/a$. As a result, the radiative recombination rate depends on $R$ only through the nanocrystal energy gap $\varepsilon_g$. Such a dependence is weak enough, as shown in Fig. 3, where $\tau^{-1}(R)$ was plotted for colloidal zinc blende CdTe nanocrystal chosen as an example. Here we used typical for CdTe and its surrounding values of $E_g = 1.5$eV, $\mu = 0.08m$, $\epsilon_0 = 2$ and set $n(\omega_g/2)\kappa(\omega_g/2) = 1$, as before. In this case the rate values vary within $\sim 10^8 \div 10^9$s$^{-1}$. It is possible to see that the single-electron radiative transition in the nanocrystal of direct-band-gap material in the whole turns out to be faster than the considered here two-electron transition in silicon nanocrystal. However, at small radii $R\sim 1$ nm these transitions have very close, and sufficiently high, rate values $\sim 10^9$ s$^{-1}$.

Phonon-assisted single-exciton creation (or annihilation) in Si nanocrystal is a slow enough process.~\cite{Hybertsen,Delerue,Moskal,AOT} One can see in Fig. 3 that its rate does not exceed $10^5$ s$^{-1}$.~\cite{AOT} Within the considered here range of nanocrystal radii this rate always remains several orders of magnitude less than the rate of the biexciton creation by a single photon.

As was demonstrated earlier both experimentally and theoretically,~\cite{Tetel,Fujii,FTT,JPCM,PRB09,Nomoto} it is possible to achieve some increase in the efficiency of the single-electron inter-band transitions in Si nanocrystals through their doping with phosphorus. Calculated recombination rate for the central donor position~\cite{PRB09} inside the nanocrystal is also shown in Fig. 3. As seen, some increase of the rate, indeed, takes place, but is insufficient to exceed the rate of the biexciton creation.

Surface modification can also influence radiative single-electron transitions in Si nanocrystals, especially, in the case of their small sizes. In particular, experimental and theoretical investigations of small (less than 1 nm in diameter) Si nanocrystals with various ligands~\cite{Light,Yang,Galar,Reyes,Momeni,Sinel} or CH$_3$-coating~\cite{Light,Kusova,Kusova1,Poddubny,JL} demonstrate transitions with rates varying within 10$^6$ -- 10$^8$ s$^{-1}$ (not shown in the figure).

Thus, the biexciton generation by a single photon in Si nanocrystals turns out to be faster than the similar process for an exciton, even in the presence of some factors promoting the exciton creation.

\section{Conclusion}

Thus, we have calculated the rate of direct biexciton generation initiated by a photon in Si nanocrystal. Within the strong quantum confinement regime the dependence of $\tau^{-1}$ on the nanocrystal radius is close to $R^{-6}$.  The rate substantially increases as the nanocrystal radius decreases and becomes of the same order of magnitude as the rate of single-exciton generation in nanocrystals of direct-band-gap materials. It is a consequence of effective enhancement of the Coulomb interaction (parameter $U_0$ in Eq. (39)) in the nanocrystal which, in turn, leads to increasing overlap of the two-electron wave functions in the ${\bf k}$-space. We can conclude that at small sizes silicon crystallites may be optically active, similarly to typical direct-band-gap semiconductors and semiconductor nanocrystals, but at higher photon energies equal to twice the nanocrystal energy gap.

\section*{Acknowledgments}

The work was supported by the Russian Science Foundation (RSF) through the project No 23-22-00275.

\section*{Data availability}

The data that supports the findings of this study are available from the corresponding author upon reasonable request through E-mail: vab3691@yahoo.com


\begin{thebibliography}{100}

\bibitem{Klimov3}
V.I. Klimov, Annu. Rev. Condens. Matter Phys. {\bf 5}, 285 (2014).

\bibitem{Marri}
I. Marri, S. Ossicini, Nanoscale {\bf 13}, 12119 (2021).

\bibitem{Melnychuk}
C. Melnychuk, P. Guyot-Sionnest, Chem. Rev. {\bf 121}, 2325 (2021).

\bibitem{Klimov1}
R.D. Schaller, V.M. Agranovich, V.I. Klimov, Nat. Phys. {\bf 1}, 189 (2005).

\bibitem{Klimov2}
V.I. Rupasov, V.I. Klimov, Phys. Rev. B {\bf 76}, 125321 (2007).

\bibitem{Beard}
M.C. Beard, K.P. Knutsen, P. Yu, J.M. Luther, Q. Song, W.K. Metzger, R.J. Ellingson, A.J. Nozik, Nano Lett. {\bf 7}, 2506 (2007).

\bibitem{Gordi}
M. Gordi, H. Ramezani, M.K. Moravvej-Farshi, J. Phys. Chem. C {\bf 121}, 6374 (2017).

\bibitem{Hyeon}
K. Hyeon-Deuk, O.V. Prezhdo, ACS Nano {\bf 6}, 1239 (2012).

\bibitem{Deuk}
K. Hyeon-Deuk, O.V. Prezhdo, J. Phys.: Condens. Matter {\bf 24}, 363201 (2012).

\bibitem{Prezhdo}
O.V. Prezhdo, Chem. Phys. Lett. {\bf 460}, 1 (2008)

\bibitem{Shabaev}
A. Shabaev, Al.L. Efros, A.J. Nozik, Nano Lett. {\bf 6}, 2856 (2006).

\bibitem{Rabani}
E. Rabani, R. Baer, Chem. Phys. Lett. {\bf 496}, 227 (2010).

\bibitem{JETP}
V.A. Burdov, JETP {\bf 94}, 411 (2002).

\bibitem{PRB07}
V.A. Belyakov, V.A. Burdov, Phys. Rev. B {\bf 76}, 045335 (2007).

\bibitem{Landolt}
Landolt-B\"{o}rnstein, {\it Numerical Data and Functional Relationships in Science and Technology}, edited by O. Madelung, M. Schulz, and H. Weiss (Springer, Berlin, 1982).

\bibitem{KL1}
W. Kohn, J.M. Luttinger, Phys. Rev. {\bf 97}, 1721 (1955).

\bibitem{KL2}
W. Kohn, J.M. Luttinger, Phys. Rev. {\bf 98}, 915 (1955).

\bibitem{Gibbs}
A. Thr\"{a}nhardt, C. Ell, G. Khitrova, H.M. Gibbs, Phys. Rev. B {\bf 65}, 035327 (2002).

\bibitem{Wilk}
R.M. Wallace, G.D. Wilk, Semicond. Int. {\bf 8}, 227 (2001).

\bibitem{Cardona}
M. Cardona, F.H. Pollak, Phys. Rev. {\bf 142}, 530 (1966).

\bibitem{Hybertsen}
M.S. Hybertsen, Phys. Rev. Lett. {\bf 72}, 1514 (1994).

\bibitem{Delerue}
C. Delerue, G. Allan, M. Lannoo, Phys. Rev. B {\bf 64}, 193402 (2001).

\bibitem{Moskal}
A.S. Moskalenko, J. Berakdar, A.A. Prokofiev, I.N. Yassievich, Phys. Rev. B {\bf 76}, 085427 (2007).

\bibitem{AOT}
V.A. Belyakov, V.A. Burdov, R. Lockwood, A. Meldrum, Adv. Opt. Tech. {\bf 2008}, 279502 (2008).

\bibitem{Tetel}
D.I. Tetelbaum, I.A. Karpovich, M.V. Stepikhova, V.G. Shengurov, K.A. Markov, O.N. Gorshkov, Surf. Invest. X-Ray Synchrotron Neutron Tech. {\bf 14}, 601 (1998).

\bibitem{Fujii}
M. Fujii, A. Mimura, S. Hayashi, K. Yamamoto, Appl. Phys. Lett. {\bf 75}, 184 (1999).

\bibitem{FTT}
D.I. Tetelbaum, O.N. Gorshkov, V.A. Burdov, S.A. Trushin, A.N. Mikhaylov, D.M. Gaponova, S.V. Morozov, A.I. Kovalev, Phys. Solid State {\bf 46}, 17 (2004).

\bibitem{JPCM}
V.A. Belyakov, A.I. Belov, A.N. Mikhaylov, D.I. Tetelbaum, V.A. Burdov, J. Phys.: Condens. Matter {\bf 21}, 045803 (2009).

\bibitem{PRB09}
V.A. Belyakov, V.A. Burdov, Phys. Rev. B {\bf 79}, 035302 (2009).

\bibitem{Nomoto}
K. Nomoto, T.C.-J. Yang, A.V. Ceguerra, T. Zhang, Z. Lin, A. Breen, L. Wu, B. Puthen-Veettil, X. Jia, G. Conibeer, I. Perez-Wurfl, S.P. Ringer, J. Appl. Phys. {\bf 122}, 025102 (2017).

\bibitem{Light}
K. Dohnalova, A.N. Poddubny, A.A. Prokofiev, W.D.A.M. de Boer, C.P. Umesh, J.M.J. Paulusse, H. Zuilhof, T. Gregorkiewicz, Light: Sci. Appl. {\bf 2}, e47 (2013).

\bibitem{Yang}
Z. Yang, G.B. De los Reyes, L.V. Titova, I. Sychugov, M. Dasog, J. Linnros, F.A. Hegmann, J.G.C. Veinot, ACS Photonics {\bf 2}, 595 (2015).

\bibitem{Galar}
P. Galar, T. Popelar, J. Khun, I. Matulkova, I. Nemec, K. Dohnalova Newell, A. Michalcova, V. Scholtz, K. Kusova, Faraday Discuss. {\bf 222}, 240 (2020).

\bibitem{Reyes}
G.B. De los Reyes, M. Dasog, M.X. Na, L.V. Titova, J.G.C. Veinot, F.A. Hegmann, Phys. Chem. Chem. Phys. {\bf 17}, 30125 (2015).

\bibitem{Momeni}
A. Momeni, M.H. Mahdieh, J. Lumin. {\bf 176}, 136 (2016).

\bibitem{Sinel}
R. Sinelnikov, M. Dasog, J. Beamish, A. Meldrum, J.G.C. Veinot, ACS Photonics {\bf 4}, 1920 (2017).

\bibitem{Kusova}
K. Kusova, O. Cibulka, K. Dohnalova, I. Pelant, J. Valenta, A. Fucikova, K. Zidek, J. Lang, J. Englich, P. Matejka, P. Stepanek, S. Bakardjieva, ACS Nano {\bf 4}, 4495 (2010).

\bibitem{Kusova1}
K. Kusova, J. Non-Crystal. Solids {\bf 358}, 2130 (2012).

\bibitem{Poddubny}
A.N. Poddubny, K. Dohnalova, Phys. Rev. B {\bf 90}, 245439 (2014).

\bibitem{JL}
N.V. Derbenyova, A.A. Konakov, V.A. Burdov, J. Lumin. {\bf 233}, 117904 (2021).

\end{thebibliography}
\end{document}